\documentclass{article}
\usepackage{spconf,amsmath,graphicx}
\usepackage{subfigure}
\usepackage{amssymb, amsfonts}
\usepackage{booktabs}  
\usepackage{multicol}  
\usepackage{multirow} 
\usepackage{pgfplots, pgfplotstable}
\usepackage{url}
\title{AVQVC: One-shot Voice Conversion
by Vector Quantization with applying contrastive learning}
\name{Huaizhen Tang$^{1,2}$, Xulong Zhang$^1$, Jianzong Wang$^{1*}$, Ning Cheng$^1$, Jing Xiao$^1$
\thanks{$^*$ Corresponding author: Jianzong Wang, (jzwang@188.com).}}
\address{\small $^1$Ping An Technology (Shenzhen) Co., Ltd.\\
\small $^2$University of Science and Technology of China}

\begin{document}
%\ninept
%
\maketitle
\begin{abstract}
Voice Conversion(VC) refers to changing the timbre of a speech while retaining the discourse content. Recently, many works have focused on disentangle-based learning techniques to separate the timbre and the linguistic content information from a speech signal. Once successful, voice conversion will be feasible and straightforward. This paper proposed a novel one-shot voice conversion framework based on vector quantization voice conversion (VQVC) and AutoVC, called \textbf{AVQVC}. A new training method is applied to VQVC to separate content and timbre information from speech more effectively. The result shows that this approach has better performance than VQVC in separating content and timbre to improve the sound quality of generated speech.
\end{abstract}
\section{Introduction}
\label{sec:intro}

Voice conversion (VC) aims to convert an utterance of a source speaker to another utterance of a target person by keeping the content in the original utterance and replacing it with the vocal features from the target speaker. Recently, considerable effort was spent on the topic of VC. Up to now, many methods have been applied in VC successfully\cite{GMM,DNN2, VAE,cross-domainVAE}. Among them, some approaches of VC require training a model for each paired speaker using parallel corpora, which limits the ability to produce natural speech for a target speaker without enough pair source-target data. To address this problem, more and more attention has focused on non-parallel VC systems.

Recently, many Generative Adversarial Networks (GAN) \cite{GAN, vaw-gan, cyclegan, stargan-vc, starganvc2} have been successfully applied to non-parallel VC. These GAN-based models jointly train a generator network with a discriminator. Where adversarial loss derived from the discriminator encourages the generator outputs to build indistinguishable from real speech. Due to the cycle consistency training, GAN-based VC models can be trained with non-parallel speech datasets.

Besides, learning discrete representations of speech has also gathered much attention. Vector Quantization (VQ), an effective data compression technology, can quantify continuous data into discrete data. Previous studies have confirmed that the quantized discrete data from the input continuous speech data is closely related to the phoneme information\cite{proveVQ}. Recently, VQVC \cite{VQVC} has been proposed to learn to disentangle the content and speaker information with reconstruction loss only. Then, VQVC+ \cite{vqvc+} was also proposed to improve the conversion performance of VQVC by adding the U-Net architecture within an auto-encoder-based VC system.

There is also another line of research focus on learning continuous speech representations via predicting context information\cite{cpc1,cpc2}. VQ-wav2vec\cite{vq-wav2vec} combines this line of research with VQ to learn discrete speech representations, and Koshizuka T \textit{et al.} \cite{vq-wav2vec1} introduced pre-trained VQ-wav2vec to achieve any-to-many voice conversion. Meanwhile, some research also tries to combine VQ with other existing work. For example, VQ-VAE \cite{vq-vae} combines VQ and VAE to improve the learning ability to disentangle the content and timbre information. VQ-CPC \cite{vq-cpc} was also proposed by combining VQ and CPC.

Unfortunately, all these VQ-based models have their inherent disadvantages. For example, the training of VQVC is simple and fast enough, but the audio quantity of this method is very poor. VQVC+ improves the conversion performance of VQVC while sacrificing the simple network structure. VQ-VAE, VQ-CPC, and VQ-wav2vec have the same problem. That is, they all introduce other network structures, which makes the model very complex.

Recently, the emerging of the model called AutoVC has given us great inspiration. By applying a speaker encoder pretrained with GE2E loss~\cite{GE2E, GE2E1, tgavc}, maximizing the embedding similarity between different utterances of the same speaker, and minimizing the similarity between different speakers, AutoVC can easily get the ideal speaker embedding. To take advantage of this work, we can disentangle the content and speaker embedding more reasonably. In this way, we can improve the conversion performance of VQVC without increasing any algorithm complexity of the model. 

This paper proposed a novel voice conversion framework that combines the AutoVC and VQVC systems, named AVQVC. Specifically, we redesign the training of VQVC by learning from the idea of AutoVC to force our model to separate the linguistic and timbre information correctly. Experiment results are carried out on the VCTK datasets. Our main contributions are as follows:

\begin{itemize}
    \item We applied a new training method in VQVC to guide discrete vector to be closer to the content features while the mean difference between encoder output and discrete vector is encouraged to be more closer to the speaker information;
    \item Compared to AutoVC, we can quickly get content embedding and speaker embedding with only one codebook structure so that we do not need to introduce a pre-trained speaker encoder like AutoVC;
    % \item Compared to some VQ-based models, our proposed model obtained a significant improvement in the evaluation of the mean opinion score and conversion similarity with lower model complexity.
\end{itemize}

\section{Method}
\subsection{VQVC}

% \begin{figure}[t]
%     \centering
%     \subfigure[Training of VQVC]{
%         \label{vqvc-training}
%         \begin{minipage}[b]{1\linewidth}
%             \centering
%             \includegraphics[width=0.85\textwidth]{iclr2022/iclr2022/VQVC_training_1.jpg}
%         \end{minipage}
%     }
%     \subfigure[Conversion of VQVC]{
%          \label{vqvc-conversion}
%         \begin{minipage}[b]{1\linewidth}
%             \centering
%             \includegraphics[width=0.85\textwidth]{iclr2022/VQVC_inference_2.jpg}
%         \end{minipage}
%     }
%     \caption{Framework of VQVC. $C_x$ is the discrete code which is produced by $codebook$. $S_x$ denotes speaker embedding, and it is gengrated from the mean difference between encoder output and $C_x$}
% \end{figure}

\begin{figure}[htp]
  \begin{minipage}[b]{1\linewidth}
            \centering
            \includegraphics[width=0.95\textwidth]{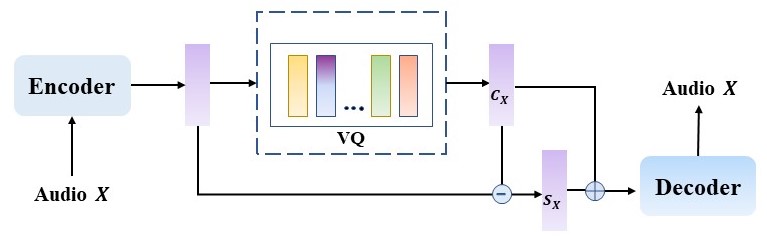}
        \end{minipage}
    \caption{Framework of VQVC. $\boldsymbol C_x$ is the discrete code which is produced by $codebook$. $\boldsymbol S_x$ denotes speaker embedding, and it is gengrated from the mean difference between encoder output and $\boldsymbol C_x$.}
    
    \label{vqvc-training}
    \vspace{-1.5em}
\end{figure}

VQVC designed a simple framework to disentangle the content embedding and speaker embedding with only one reconstruction loss. As shown in Figure \ref{vqvc-training}, the framework of VQVC contains three network structures: an encoder to extract latent features from the input speech, a learnable codebook that quantizes continuous data into discrete data, and a decoder that produces the converted speech from the content and speaker embedding. We regard the discrete data generated by Codebook as content embedding. Then, we can quickly get speaker embedding from the mean difference between encoder output and discrete data. The discrete content embedding $\boldsymbol {C_x}$ and the speaker embedding $\boldsymbol S_x$ can be derived as
\begin{flalign}
    &&
    C_{x} = VQ(enc(x)) 
    &&
    S_{x} = {E_t}[enc(x) - C_{x}]
    &&
\end{flalign}
\noindent Where we define $enc(\cdot)$ as the encoder, The expectation $E_t$ takes on the segment length on latent space, and $VQ$ was denoted as the quantization function which can quantize sequence of continuous data into the closest discrete code. Specifically, if we define $V$ as a sequence of continuous data, that is, $\boldsymbol V = v_{1}, v_{2}, ... v_{T}$. Then $VQ(\boldsymbol V)$ can be described as

\begin{align}
    & VQ(\boldsymbol{V}) = {q_{1}, q_{2}, ... q_{T}} \\
    & q_{j} = \arg\min_{q \in
    Codebook}(\|v_{j}-q\|_{2}^{2})
\end{align}

In the training phase, one utterance $x_i$ was selected randomly to do the reconstruction task. The latent-code loss function was optimized to minimize the distance between the discrete code and the continuous embedding. Besides, the self-reconstruction loss function was designed to constrain the model to find a proper balance between removing speaker information and retaining linguistic content.  They can be expressed as:

\begin{align}
   \mathcal{L}_{\text {latent}} &= \mathbb{E}[\|enc(x) -C_{x}\|_{2}^{2}] \\
   \mathcal{L}_{\text {recon }} &= \mathbb{E}[\|\hat{x}_{i \rightarrow i}-x_{i}\|_{1}^{1}] \\
    \hat{x}_{i \rightarrow i} &= Decoder(c_{x_{i}} + s_{x_{i}})
\end{align}

In conversion, two utterances from different speakers were chosen to be the source and target speech, respectively. We fed these two utterances into the trained model to get content embedding and speaker embedding, respectively. Then, the content embedding of the source speaker and the target speaker's speaker embedding were put into the decoder together to get the conversion audio from the decoder output.

% \begin{figure}[h]
%   \begin{minipage}[b]{1\linewidth}
%             \centering
%             \includegraphics[width=0.9\textwidth]{iclr2022/VQVC_inference_2.jpg}
%         \end{minipage}
%     \caption{Conversion of VQVQ/AVQVC. Note that although the training processes of VQVC and AVQVC are different, their conversion stages are completely consistent}
    
%     \label{vqvc-conversion}
% \end{figure}

\subsection{AVQVC}

\begin{figure}[t]
  \begin{minipage}[b]{1\linewidth}
            \centering
            \includegraphics[width=1\textwidth]{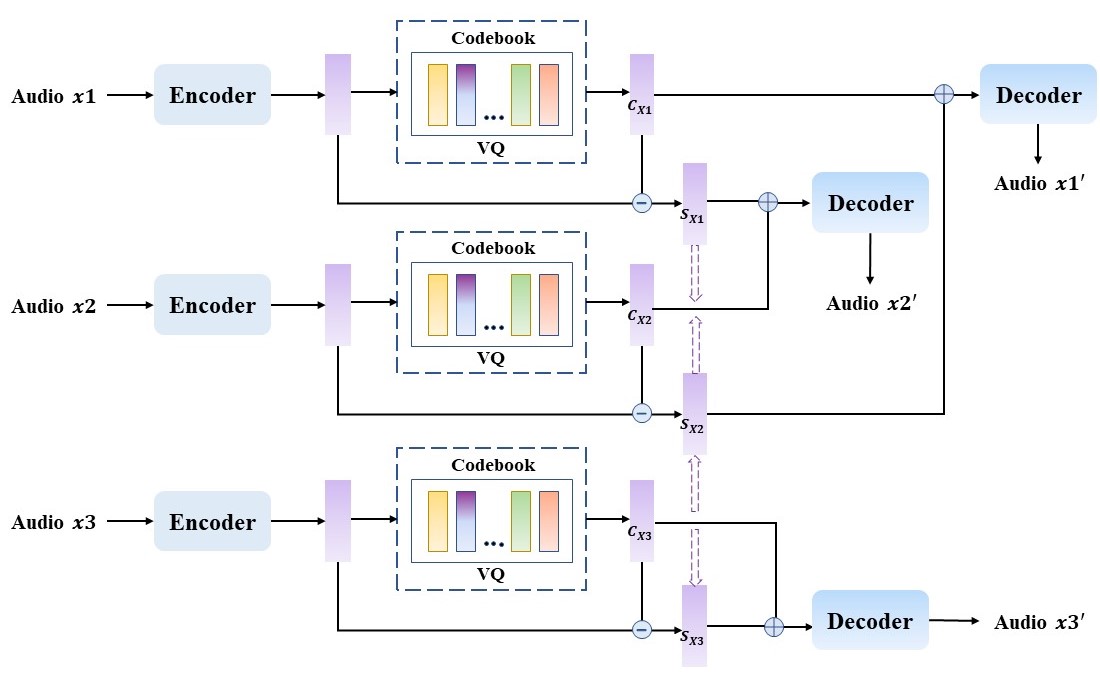}
        \end{minipage}
    \caption{Framework of AVQVC. Both $x_1$ and $x_2$ are produced by the same speaker, but their text content are different, while $x_3$ belongs to another speaker.  $\boldsymbol C_X$ is a discrete variable generated by looking up the $codebook$. And, $\boldsymbol S_X$ denotes speaker embedding, which is produced by the mean difference between encoder output and $\boldsymbol C_X$.}
    
    \label{AVQVC}
    \vspace{-1em}
\end{figure}

In this section, we will introduce the core idea of our method. As illustrated in Figure \ref{AVQVC}, the composition network structure of our model is similar to VQVC, but the training method is completely different. Specifically, the input of our model contains three sentences instead of one. And they are represented by $x_{1}$, $x_{2}$, and ${x_{3}}$ respectively. Among them, $x_{1}$ and $x_{2}$ are produced by the same speaker, but their text content is different, and $x_{3}$ is generated by another speaker. We assume that our VQ model can separate content embedding and speaker embedding correctly, which means that when $x_{1}$ and $x_{2}$ are sent to our VQ model after the same processing, their speaker embedding should be the same. Furthermore, when we put $x_{1}$ and $x_{3}$ (or $x_2$ and $x_3$) into our VQ model after the same processing, their speaker embedding should be far different. Based on this assumption, we designed a new training program.

In training phase, we input $x_{1}$, $x_{2}$, and ${x_{3}}$ into the same model at the same time, then we can easily get $\boldsymbol C_{x_{i}}$ and $\boldsymbol S_{x_{i}} $ ($i = 1,2,3$) , Then we exchange $\boldsymbol S_{x_{1}}$ and $\boldsymbol S_{x_{2}}$ to do a novel reconstruction task. Since $\boldsymbol S_{x_{1}}$ and $\boldsymbol S_{x_{2}}$ are expected to be the same, the new reconstruction speech $x'_{1}$, $x'_{2}$ should be as close to $x_{1}$, $x_{2}$ as possible. Here we still use self-reconstruction loss and latent-code loss to constrain VQ model. They can be expressed as
\begin{align}
    \mathcal{L}_{\text {recon }} &= \|x'_{1}-x_{1}\|_{1}^{1} +  \|x'_{2}-x_{2}\|_{1}^{1} + \|x'_{3}-x_{3}\|_{1}^{1}
\end{align}
\begin{multline}
    \mathcal{L}_{\text {latent}}=\|enc(x_{1}) -C_{x_{1}}\|_{2}^{2} \\ + \|enc(x_{2}) -C_{x_{2}}\|_{2}^{2}
   +\|enc(x_{3}) -C_{x_{3}}\|_{2}^{2}
\end{multline}

\noindent Where $x'_{1}$ produced by $\boldsymbol C_{x_{1}}$ and $\boldsymbol S_{x_{2}}$ , $x'_{2}$ generated from $\boldsymbol C_{x_{2}}$ and $\boldsymbol S_{x_{1}}$. And, $x'_{3}$ is special, it comes from $\boldsymbol C_{x_{3}}$ and $\boldsymbol S_{x_{3}}$, both of which are all produced by $x_3$ itself. That is
\begin{align}
    x'_{3} = Decoder(C_{x_{3}} + S_{x_{3}}).
\end{align}
In addition, we also design speaker-loss function and diff-speaker-loss function to encourage discrete data to be as close to content embedding as possible. Then, the mean difference between continuous and discrete data will naturally become closer and closer to speaker embedding. Specifically, for $\boldsymbol S_{X_{1}}$ and $\boldsymbol S_{X_{2}}$, since their speaker is the same one, we expect their speaker embeddings to be as close as possible. Similarly, because $\boldsymbol S_{X_{1}}$ and $\boldsymbol S_{X_{3}}$ (or $\boldsymbol S_{X_{2}}$ and $\boldsymbol S_{X_{3}}$ ) are produced by different speakers, we expected their speaker embeddings to be as different as possible. These two loss functions can be computed as  
\begin{align}
    &\mathcal{L}_{\text {speaker}} = \|S_{x_{2}} -S_{x_{1}}\|_{1}^{1} \\
    &\mathcal{L}_{\text {diff}} = -(\|S_{x_{2}} -S_{x_{3}}\|_{1}^{1} + \|S_{x_{1}} -S_{x_{3}}\|_{1}^{1})
\end{align}
\noindent Then the full objective function can be formulated as
\begin{align}
    L = \mathcal{L}_{\text {recon}} + \alpha \mathcal{L}_{\text {latent}} + \beta \mathcal{L}_{\text {speaker}} + \lambda \mathcal{L}_{\text {diff}}
\end{align}

The inference phase of AVQVC is completely consistent with that of VQVC. One utterance of the source speaker and the target speaker is selected to get content embedding and speaker embedding respectively. After that, we input them into the decoder, and the conversion speech is then generated.

\section{Experiments}
\label{sec:format}
\subsection{Experimental Setup}
We evaluated our proposed method on the VCTK Corpus\cite{VCTK}, which contains 46 hours of speech data produced by 109 English speakers from different countries. In our work, the entire dataset is randomly divided into three sets: 90 speaker recordings for training, 10 speaker recordings for evaluation and other 4167 recordings from 9 speakers for testing. And, the sampling rate of all recordings is 16khz, and the mel-spectrograms are computed through a short-time Fourier transform (STFT) with Hann windowing, where 1024 for FFT size, 1024 for window size and 256 for hop size. The STFT magnitude is transformed to the mel scale using 80 channel mel filter bank spanning 90 Hz to 7.6 kHz. 

We train AVQVC model with batch size of sixteen for 1M steps on one NVIDIA V100 GPU, and ADAM optimizer was used with $\beta_{1} = 0.9, \beta_{2} = 0.98$. The weights in Eq.(13) are set to $\alpha = 0.02, \beta = 0.03, \lambda = 0.02$. It is worth noting that with the increase of $\lambda$ , the training of our model becomes somewhat unstable, and when $\lambda$ is too small, it will be difficult for our model to remove the speaker information from the input speech. In fact, in order to ensure the stability of model training, when the value of $\mathcal{L}_\text{diff}$ is five times more than $\mathcal{L}_\text{recon}$, we will reduce $\lambda = 0.01$. At the same time, we will also increase $\beta = 0.05$ and change the weights of recon-loss to 2, which can encourage the model to learn codebook closer to continuous data so as to improve the converted audio quality. 

\begin{table*}[htp]
    \vspace{-2em}
  \centering
  \fontsize{8}{7}\selectfont
  \setlength{\abovecaptionskip}{0pt}%    
  \setlength{\belowcaptionskip}{10pt}%
  \caption{Comparison of different models in traditional VC and one-shot vc.}
  \label{Comparison}
    \begin{tabular}{cccccccc}
    \toprule
    \multirow{2}{*}{\textbf{Methods}}&
    \multicolumn{3}{c}{\textbf{Traditional VC}}&\multicolumn{3}{c}{\textbf{ One-Shot VC}}\cr
    \cmidrule(lr){2-4} \cmidrule(lr){5-7}
    & MCD & MOS & VSS & MCD & MOS & VSS & MODEL-SIZE\cr
    \midrule
    VQVC & 8.16 $\pm$ 0.31
    & 2.28 $\pm$ 0.99 
    & 3.47 $\pm$ 0.82 & 8.12 $\pm$ 0.14 
    & 2.06 $\pm$ 0.84 & 2.97 $\pm$ 0.75
    & \textbf{5.71M}  \cr
    VQVC+ & 7.08 $\pm$ 0.22 & 3.31 $\pm$ 0.90 
    & 3.42 $\pm$ 0.85 & 8.41 $\pm$ 0.08 
    & 2.75 $\pm$ 0.84 & 3.11 $\pm$ 0.88
    & 388M  \cr
    AutoVC& \textbf{4.34 $\pm$ 0.12} & \textbf{3.81 $\pm$ 1.14}
    & 3.45 $\pm$ 0.76 & 7.66 $\pm$ 0.17 
    & 2.61 $\pm$ 0.73 & 2.91 $\pm$ 0.72
    & 339M  \cr
    StarGAN-VC2& 6.28 $\pm$ 0.09 & 3.45 $\pm$ 1.01
    & 3.59 $\pm$ 0.87 & — & — & — & 56.45M  \cr
    \midrule
    \textbf{AVQVC(512)} &5.19 $\pm$ 0.29 &3.57 $\pm$ 0.91
    &\textbf{3.70 $\pm$ 0.71} &\textbf{5.04 $\pm$ 0.13} 
    &\textbf{3.20 $\pm$ 0.91} &\textbf{3.29 $\pm$ 0.64}
    &5.77M  \cr
    \bottomrule
    \end{tabular}
    \vspace{-1.2em}
\end{table*}

\subsection{Comparison}

Here we will evaluate the performance of our method in traditional VC tasks and one-shot VC tasks. Specifically, traditional VC means that both the selected source speaker and the target speaker already exist in the training set. And one-shot VC refers to a new task only needs one utterance from the source speaker and target speaker, and both these two speakers do not need to appear in the training set. To compare the performance of VC between our method and other previous works, the Mel-Cepstral Distortion(MCD) between converted speech and the ground truth target speech as our objective evaluation and two subjective evaluation methods were also introduced. One is the mean opinion score(MOS) test, which is used to evaluate the quality of converted speech. Specifically, we invited 12 humans (seven males and five females) participants to evaluate the quality of some converted speech generated from different models. After hearing each speech, the subjects should choose a score from 1-5 points of the naturalness of the converted speech. The higher the score, the better they think the audio quality of the speech. The other is the voice similarity score(VSS) test, which measures how similar the timbre of the converted voice is to that of the ground truth. And its scoring mechanism is the same as that of MOS. The higher the score, the more similar the tone between the converted speech and the target speech. AutoVC, VQVC, VQVC+, StarGAN-VC2 were chosen as baselines. And the result shows in table \ref{Comparison}. 

% \begin{table}[h]
%     \centering
%     \label{tvc}
%     \caption{Comparison of different models in traditional VC and one-shot vc}
% 	\begin{tabular}{  c | c | c | c | }
% 		\hline
% 		Methods 	& MCD 	& MOS	 & SMOS \\ 
% 		\hline
% 		VQVC 	& 11C		& 22C		 & A  \\ 
% 		\hline 
% 		VQVC+ & 9C 		& 19C		 & B  \\ 
% 		\hline
% 		VQ-VAE 	& 11C		& 22C		 & A  \\ 
% 		\hline
% 		AutoVC 	& 11C		& 22C		 & A  \\ 
% 		\hline
% 		AVQVC 	& 11C		& 22C		 & A  \\ 
% 		\hline 
% 	\end{tabular}
% \end{table}
The result shows that with the new training method, the quality of audio converted by VQVC has been significantly improved. Compared with other VQ-based models, the speech quality generated by our method is equivalent and slightly worse than that of AutoVC, but in VSS Test, our performance is better than AutoVC. Most importantly, compared with VQVC, our method only adds few parameters, which makes our model much simpler and faster than AutoVC, StarGAN-VC2, or other VQ-based models.

% To further evaluate the conversion effect of different models, we extract F0 from converted speech and ground truth speech, and we plot them in Figure \ref{f0}.

% \begin{figure}[h]
%   \begin{minipage}[b]{1\linewidth}
%             \centering
%             \includegraphics[width=0.8\textwidth]{iclr2022/icassp_2021_template/F0_comprasion.png}
%         \end{minipage}
%     \caption{Comparison of F0.}
    
%     \label{f0}
% \end{figure}

In addition, by applying a well-known open-source speech detection toolkit, \textit{Resemblyzer} (\url{https://github.com/resemble-ai/Resemblyzer}), we conduct a fake speech detection test to compare the quality and similarity of the converted speeches from different models respectively against ground truth reference audio. The higher the score, the better the quality and similarity of the converted speech. The results are shown in Table~\ref{table:1}.

\begin{table}[htbp]
    \vspace{-1em}
   \centering
   \caption{Comparison of different methods in traditional VC tasks.} 
%   $\boldsymbol{C}$:Channel dimensions of the content embedding;\\
%   $\boldsymbol{S}$: Channel dimensions of the speaker embedding.}
    \label{table:1}
    \begin{tabular}{l c}
     \hline
     Method & Score \\
     \hline
     \textbf{AVQVC} &\textbf{0.72 $\pm$ 0.24} \\
     AutoVC    &0.70 $\pm$ 0.43 \\
     VQVC+     &0.63 $\pm$ 0.31 \\
     \hline 
    \end{tabular}
    \vspace{-0.5em}
\end{table}

The result shown in Table~\ref{table:1} shows that the converted speech produced by AVQVC is more similar to the ground truth than that of AutoVC and VQVC+. It indicates AVQVC outperforms AutoVC and VQVC+ in the VC tasks.

Furthermore, we also evaluate the performance of different models in the one-shot VC task. Since StarGAN-VC2 can not achieve the VC task for unseen speakers, AutoVC, VQVC, and VQVC+ were chosen as our baseline models. The results in table \ref{Comparison} show that with only one utterance of unseen speakers, the performance of AutoVC is greatly reduced. Previous studies have reported this phenomenon~\cite{proveAUTOVC}. While in the same case, our method still has a good performance, indicating that our model has strong adaptability under both conditions.

%  The result shows that with only reconstruction loss, it is not very easy for VQVC network to learn to find a proper balance between removing the speaker information and preserving the linguistic content.

\subsection{Codebook Size}

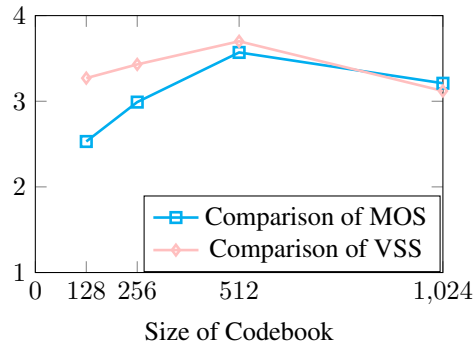
\begin{figure}[htp]
    \centering
    \begin{tikzpicture}
    \begin{axis}[
    height=5cm,
    width=7cm,
    xlabel=Size of Codebook,
    % ylabel=Score,
    xmin=0,
    xmax=1024,
    ymin=1,
    ymax=4,
    xtick = {0,128, 256, 512, 1024},
    legend style={at={(0.63,0.3)},anchor=north}
    % ytick pos=left
    ]
    \addplot[color = cyan, line width = 1, mark = square] coordinates{
    (128,2.53)
    (256,2.99)
    (512,3.57)
    (1024,3.21)
    };
    \addlegendentry{Comparison of MOS}
    \addplot[color = pink, line width = 1, mark = diamond] coordinates {
    (128,3.27)
    (256,3.43)
    (512,3.7)
    (1024,3.12)
    };
    \addlegendentry{Comparison of VSS}
    \end{axis}
    \end{tikzpicture}
    \caption{Performance comparison of MOS and VSS under different Codebook sizes}
    \label{fig:Comparison}
    \vspace{-0.4em}
\end{figure}

 In addition, we also focused on the choice of codebook size. In VQVC, 256 is selected as an appropriate number of discrete codes. And in our work, since we add some new loss functions to constraint the codebook, we don't need to worry that using a large codebook size will include speaker information, so we conduct many comparison experiments to find the proper number of codebook sizes. Finally, 512 is selected. And we will show the MOS and VSS results under different codebook sizes in Figure \ref{fig:Comparison}.
 
% \pgfplotsset{width=7cm,compat=1.13} % 图片绘制的宽度是7cm,使用的pgfplots版本为1.13
% \begin{tikzpicture}
% \begin{axis}[legend pos=in south east] % 将图例放在图外，位于图的东北角
% \addplot 
% table                               % 绘制原始数据的折线图
% {           		                % X，Y的原始数据
%  X Y
%  128 3
%  256 4
%  512 5
%  1024 5 
% };
% \addplot
% table[y={create col/linear regression={y=Y}}] % 对输入的数据作线性回归
% {   				
%  X Y
%  128 1
%  256 2
%  512 3
%  1024 4 
% };
% \addlegendentry{Comparison of MOS}          % 给第一个图像添加图例，即原始函数y(x)
% \addlegendentry{                 % 给第二个图像添加图例，即线性回归结果a*x+b
% $\pgfmathprintnumber{\pgfplotstableregressiona} \cdot x
% \pgfmathprintnumber[print sign]{\pgfplotstableregressionb}$}
% \end{axis}
% \end{tikzpicture} 

As illustrated in Figure~\ref{fig:Comparison}, when the size of the codebook is 512, the value of MOS and VSS are all significantly improved—noted that when codebook size exceeds 512, the performance of our method also decreases. We hypothesize that when the codebook's size is too large, the discrete vector will inevitably contain some timbre information, the loss function we designed will lose its constraint and our model will finally degenerate into the original VQVC.   
% \begin{tikzpicture}
% \begin{axis}[
% height=7.5cm,
% width=9cm,
% xlabel=Size of Codebook,
% % ylabel=Score,
% xmin=0,
% xmax=1024,
% ymin=1,
% ymax=5,
% % xtick ={0, 128, 256, 384, 512, 640, 768, 896, 1024},
% xtick = {0,128, 256, 512, 1024},
% legend pos= north west
% % ytick pos=left
% ]
% \addplot coordinates {
% (128,3)
% (256,3.3)
% (512,4)
% (1024,4)
% };
% \addlegendentry{Comparison of Mos}
% \addplot coordinates {
% (128,1)
% (256,3)
% (512,2)
% (1024,4)
% };
% \addlegendentry{Comparison of VSS}
% \end{axis}
% \end{tikzpicture}

\section{conclusion}
In this paper, we proposed a new model to apply a new training method to VQVC to improve the performance of voice conversion.  We conducted both objective and subjective evaluations to evaluate the performance of our method in different VC tasks. The result shows that with the new training method applied to VQVC, the quality of converted audio speech and the similarity with the target voice has significantly improved. 

\section{Acknowledgement}
This paper is supported by the Key Research and Development Program of Guangdong Province No. 2021B0101400003 and the National Key Research and Development Program of China under grant No. 2018YFB0204403.

\bibliographystyle{IEEEbib}
\bibliography{refs}

\end{document}